\documentclass[traditabstract]{aa}
\usepackage[utf8]{inputenc}
\usepackage{amsmath}
\usepackage{amssymb}
\usepackage{microtype}
\usepackage{graphicx}
\usepackage{txfonts}
\usepackage{color}
\usepackage{multirow}
\usepackage{xspace}
\bibpunct{(}{)}{;}{a}{}{,} 
\newcommand{\src}{\object{CXOU~J160103.1-513353}\xspace}

\usepackage{natbib}

\begin{document}

\title{\src: another CCO with a carbon atmosphere?}
\author{V.\,Doroshenko\inst{1}, V.\,Suleimanov\inst{1,2,3}, A.\,Santangelo\inst{1}}  

\institute{Institut für Astronomie und Astrophysik, Sand 1, 72076 Tübingen, Germany\and
Space Research Institute of the Russian Academy of Sciences, Profsoyuznaya Str. 84/32, Moscow 117997, Russia\and
Kazan (Volga region) Federal University, Kremlevskaja str., 18, Kazan 420008, Russia}

\abstract{We report on the analysis of XMM-Newton observations of the central compact
object \src located in the center of the non-thermally emitting supernova
remnant (SNR) G330.2+1.0. The X-ray spectrum of the source is well described
with either single-component carbon or two-component hydrogen atmosphere
models. In the latter case, the observed spectrum is dominated by the emission
from a hot component with a temperature $\sim3.9$\,MK, corresponding to the
emission from a hotspot occupying $\sim1$\% of the stellar surface (assuming a
neutron star with mass $M=1.5M_\odot$, radius of 12\,km, and distance of
$\sim5$\,kpc as determined for the SNR). The statistics of the spectra and
obtained upper limits on the pulsation amplitude expected for a rotating
neutron star with hot spots do not allow us to unambiguously distinguish between
these two scenarios. We discuss, however, that while the non-detection
of the pulsations can be explained by the unfortunate orientation in \src,
this is not the case when the entire sample of similar objects is considered.
We therefore conclude that the carbon atmosphere scenario is more plausible.
}

\keywords{Stars: neutron, X-rays: stars, Stars: atmospheres}
\authorrunning{V. Doroshenko et al.}
\maketitle

\section{Introduction} The central compact object (CCO) \src was discovered
with \emph{Chandra} at the center of the supernova remnant (SNR) G330.2+1 by
\citet{Park06}. The origin of the source was established based on the absence
of counterparts in other wavebands and on the observed blackbody like
spectrum with $kT\sim0.5$\,keV typical for other CCOs.
\citet{Park06} also claimed the tentative detection of pulsations from the source with a period of
$\sim7.5$\,s and an amplitude of $\sim30$\,\%, which would clearly establish
the object as a neutron star. However, follow-up observations with
\emph{XMM-Newton} refuted this detection \citep{Park09}. Poor counting
statistics only allowed to place an upper limit on the pulsed fraction $PF=(F_{\rm max}-F_{\rm min})/(F_{\rm max}+F_{\rm min})$ of
$\sim40$\,\%. This is significantly higher than that observed in other CCOs, which means that
observations were likely not sensitive enough to detect pulsations
from the source.

The source, however, is expected to pulsate. While the observed X-ray
spectrum could be described as a uniformly emitting neutron star hydrogen atmosphere, this
results in unreasonably large distances $\ge24$\,kpc for standard neutron star
parameters \citep{Park09}. The actual distance to the SNR was estimated by
\citet{McClure_distance} at $\sim 5$\,kpc based on the observed H~I
absorption in radio, which implies a significantly smaller emission region size.
\citet{Park09} therefore concluded that the observed emission
predominantly arises from hotspots with radii of 0.3-2\,km, while the remaining surface is significantly cooler and does not contribute much to the
observed flux. This scenario was first proposed by \citet{Pavlov2000} for the
CCO in Cas A, and it implies that the source must pulsate, which has not been observed so far.

In addition to the limited sensitivity of the observations, the
non-detection of the pulsations might be explained by geometrical
considerations, that is, alignment of the spin axis of the neutron star with the
magnetic field or with the line of sight \citep{Suleimanov17}, especially given the
existing modest constraints on the observed pulsed fraction. Alternatively, the
lack of pulsations can be explained if the observed X-ray emission does arise from the entire surface of the neutron star whose atmosphere is composed
of heavier elements such as carbon \citep{Ho09,Klochkov13,
Klochkov14,Klochkov16}. This is an intriguing option as pure hydrogen
atmospheres are commonly assumed to interpret thermally emitting
neutron star observations \citep{Pavlov04}. Such observations are considered one of
the prime probes for the equation of state of matter under supra-nuclear
densities, and thus are of fundamental importance \citep{Lattimer07,Lattimer16}.

The possibility that the atmosphere consists of heavier elements despite
presumably fast stratification complicates the interpretation of the
observations, so that additional observational information such as cooling of the
neutron star must be considered in this case \citep{Ofengeim15}. On the other
hand, detection and observed amplitude of the pulsations must also be taken
into consideration. For instance, \citet{Bogdanov14} demonstrated that the
spectrum of another thermally emitting neutron star PSR~J1852+0040 in SNR Kes~79 can be well
described with a single-temperature carbon atmosphere. However, strong
pulsations are observed in this case, which suggest that caution must be taken
when making any conclusions based on the observed spectrum alone.

We here discuss \src in this context. Using the new 150\,ks
\emph{XMM-Newton} observation of the source, we obtain improved X-ray spectra
and upper limits on the fraction of pulsed emission. We conclude that
observations are fully consistent with the carbon atmosphere scenario. On the other
hand, the available data do not allow us to univocally exclude the two-temperature
hydrogen atmosphere model based on non-detection of the pulsations, so
that additional observations might be required. 
Nevertheless, 
a rather particular geometry is required to explain the absence of 
pulsations in this and several other non-pulsating CCOs, which suggests that the a
single carbon atmosphere model might still be preferred.

\section{Data analysis and results} The source has been observed with
\emph{Chandra} \citep{Park06}, \emph{ASCA} \citep{Torii06}, and several times
with \emph{XMM-Newton}. Here we use data from the two longest \emph{XMM}
observations: obsid. 0500300101 (68\,ks) and 0742050101 (140\,ks). The first
observation is significantly affected by in-orbit background, and has previously
been used by \citet{Park09} to characterize the spectrum of the CCO, while the
second observation has only been used so far to study the extended emission
from the SNR by \citet{Williams18}, who confirmed the nonthermal origin of the extended
emission and strong absorption in the direction of the source with an equivalent
column of $\sim2-3\times10^{22}$\,cm$^{-2}$.

The data reduction was carried out according to the XMM user guide using the
\emph{XMM~SAS} version 16.1.0 and a current set of calibration files. Both
pointings were contaminated by soft proton background to some extent, so that after
the standard screening, the effective exposures were reduced to $\sim35$\,ks and
93/110\,ks (PN/MOS) for the first and second observations, respectively. We only
used EPIC MOS data for the first observation since the PN was operated in
small-window mode, with some extended emission falling within the storage area
and thus increasing the background. Together with uncertainties in calibration
in small-window mode, this made this pointing not very useful for spectral analysis,
and a timing analysis has previously been published by \citet{Park06}.

The spectrum of the source and events for timing analysis were extracted from a
circle with a radius of $28^{\prime\prime}$ centered on the coordinates reported
by \citet{Park06}. The extraction radius was selected to optimize the signal-to-noise ratio with the help of the \emph{eregionanalyse} task. The background was
extracted from two circles with the same radius adjacent to the source with the
same \emph{RAWY} coordinate to ensure that similar low-energy noise was
subtracted for EPIC
PN\footnote{http://xmm2.esac.esa.int/docs/documents/CAL-TN-0018-2-12.pdf}. Two
background regions were selected to account for possible variations in
brightness of the extended emission. We verified that results were not
significantly affected when the background was extracted from an annulus around the
source (which is, however, not recommended, as detailed in the same calibration
note). The same source background regions were also used for the MOS cameras.

\paragraph{Spectral analysis} Source spectra were extracted for two
observations and three cameras and were simultaneously fit together after binning
to contain at least 25 counts/bin. As discussed by \citet{Park06,Park09}, a
single blackbody or hydrogen atmosphere model results in an unrealistically large
distance to the source. Therefore, we only considered the two-component,
non-magnetic hydrogen and carbon atmosphere models
\citep{Suleimanov14,Suleimanov17}.
To fit the spectra, we used \emph{Xspec} version 12.9.1p, where both models are
implemented (\emph{hatm} and \emph{carbatm,} respectively). Additionally, we
included the \emph{TBabs} absorption component and cross-normalization
constants to account for interstellar absorption \citep{Wilms00} and effective
area discrepancy between individual instruments. In both cases we assumed
$M=1.5M_\odot$ and $R=12$\,km for the neutron star, and a distance to the source of
4.9\,kpc \citep{McClure_distance}. 

For the two-component fit the normalizations of the components were linked to
each other through an additional constant $0\le\delta\le1$ defining the fraction
of the surface area occupied by the hotspots \citep{Suleimanov17}. In particular,
we assumed that the model consists of two atmosphere model spectra of different temperatures
with a free ratio of their emitting areas:
$$
{f^{'}}_E=\frac{R^2}{d^2(1+z)}(\delta F_{E^{'}}(T_1)+(1-\delta)F_{E^{'}}(T_2)),
$$
where the emitted $E^{'}$ and the observed energies $E$ of the photons are connected
by the relation $E^{'}=E(1+z),$ where $z=(1-R_s/R)^{-1/2}-1$ and
$R_s=2GM/c^2$  are the gravitational redshift and Schwarzschild radius, and
$d$ is the assumed distance to the source.

\begin{figure}[t]
    \centering
        \includegraphics[width=\columnwidth]{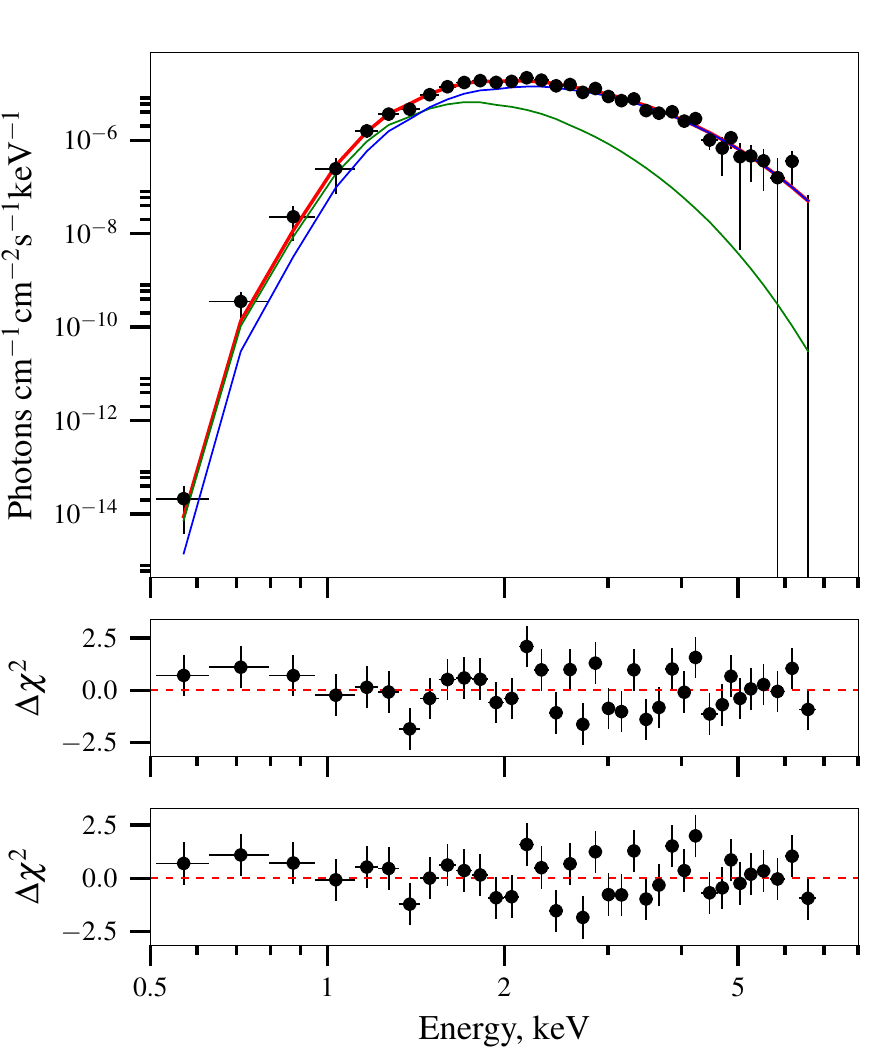}
    \caption{Best-fit unfolded spectrum (top panel) and residuals for the two-component hydrogen atmosphere (middle panel)
    and carbon models (bottom panel).
    Only the EPIC PN spectrum from observation 0742050101 is shown for clarity. The contribution of
    the cold and hot components to the two-component fit is shown with thin lines. The
    total model flux is indistinguishable for the two models and is shown with the thick red line.}
    \label{fig:spe}
\end{figure}

The best-fit results and a representative spectrum are presented in
Table~\ref{tab:fit} and Fig.~\ref{fig:spe}. Both models provide an adequate
description of the data, and neither is statistically preferred.

It is interesting to note that both models require a significantly higher
absorption column than that derived by \citet{Williams18} for
extended SNR emission ($N_H\le3.13\times10^{22}$\,atoms\,cm$^{-2}$). This mismatch
might point to problems with the modeling of the extended emission spectrum or 
to an overestimation of the soft X-ray flux of the neutron star by atmosphere models, or
it might indicate additional absorption specific to neutron star. We note that \src is not
the only object where the CCO emission appears to be more absorbed than
the extended emission of the remnant. A comparison of the absorption
columns reported by \citet{Klochkov14} for CCO in SNR G353.6-0.7 and by
\citet{Doroshenko17} for the extended emission around the neutron star also reveals
an enhanced absorption for the compact object.

The large difference in temperature of the two components and the high absorption
column in the two-component fit also implies that most of the flux from the
cold surface of the neutron star is not observed. The emission from the
hotspots therefore dominates the observed flux, which contributes $\sim82$\%. This
is in line with the analysis results of the first observation by
\citet{Park09}, who found that the soft component is not statistically
required by the fit.

\begin{figure}[t]
    \centering
        \includegraphics[width=\columnwidth]{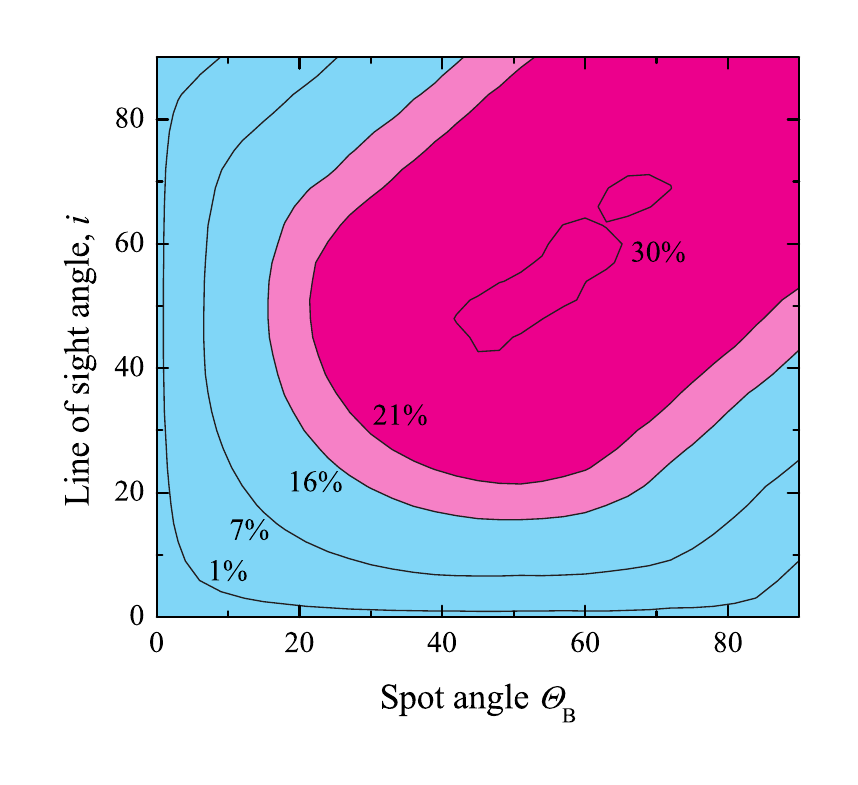}
    \caption{Contours of constant PF in the $\theta_{\mathrm B}-i$ plane. The permitted region with PF$\le16-21$\%
    (depending on the considered period range) is shown in blue. }
    \label{fig:map}
\end{figure}

\begin{table}
    \begin{center}
    \begin{tabular}{lcc}
        Parameter & \emph{hatm} & \emph{carbatm} \\
        \hline
        $N_{\rm H}$\,10$^{22}$\,, atoms\,cm$^{-2}$ &  4.8(4) & 5.0(1) \\
        $kT_1$\,,keV & 1.4(2) & 1.73(1) \\
        $kT_2$\,,keV & 3.9(2) & \\
        $\delta$ & 0.0098(4) & \\
        \hline
        $\chi^2$/dof & 107.6/114 & 115.0/116\\
    \end{tabular}
    \end{center}
    \caption{Best-fit parameters for the two-component hydrogen and carbon atmosphere models assuming
    a distance fixed to 4.9\,kpc. Uncertainties are quoted at the $1\sigma$ confidence level}
    \label{tab:fit}
\end{table}

\paragraph{Timing analysis} We also exploited the additional exposure provided by
the new data to improve the $\sim40$\% upper limit on the pulsed fraction
obtained by \citet{Park09} based on the first \emph{XMM} observation. Other CCOs have short periods, therefore the important constraint here was
provided by EPIC PN data, which offered 5.7\,ms time resolution in small-window
mode. Unfortunately, the obtained limit is not particularly constraining because of
the high background level ($\sim30$\%), and we were unable to significantly
improve it. On the other hand, EPIC MOS operated in full-frame mode and thus is
only sensitive for periods above $\sim5.2$\,s, which is not enough
by far to probe the $\sim100$\,ms periods typical for CCOs \citep{Gotthelf13}.

The time resolution of EPIC PN in full-frame mode used for the second
observation is $\sim73$\,ms, which is not sufficient either to probe
timescales quite as short as this. However, pulsations with periods longer $\sim147$\,ms can be probed,
and at least one CCO has a period in this range. The observation is also longer by
a factor of three than the first one and has a lower level of in-orbit
background, which significantly increases the sensitivity to the pulsations in the
accessible range of periods and thus allows us to improve on the upper limits obtained by
\citet{Park09}. The upper limits for the MOS data can likewise clearly be improved,
so that we considered a period range of 147\,ms to 100\,s for EPIC PN data and 5.2-100\,s for
the combined data set including data from both cameras.

To search for the pulsations and estimate upper limits on the pulsed fraction,
we used the same methods as \citet{Doroshenko15}. First, we optimized the energy
range and extraction radius for the source in order to obtain the highest signal-to-noise ratio. The energy range was chosen to minimize contribution of the
background (mostly extended emission from the SNR). Then the optimal
extraction radius of $28^{\prime\prime}$ was estimated with
the \emph{eregionanalyse} task for the obtained energy range of 1.7 to 4.2\,keV.
Coincidentally, this energy range also mostly excludes photons from the cool
component in the two-component atmosphere model, which means that the intrinsic pulsations must also be
the strongest. After the extraction, the photon arrival times were corrected for using
the \emph{barycen} task, and a search for pulsations was carried out using the
H-test of \citet{Jager89}. Similarly to \citet{Park06}, we oversampled the search
frequency grid by factor of ten to ensure that no peaks were missed. In line
with the previous findings, no significant pulsed signal was found.

To estimate the upper limit on the pulsed fraction, we followed the approach by
\citet{Brazier94,Doroshenko15} and repeated the procedure separately for the
combined MOS/PN and PN data. The strictest upper limits were obtained by
combining both observations (assuming that the period of the source did not
change), although we also carried out a timing analysis for individual
observations. For periods in the range of 5.2\,s to 100\,s, 4520 photons
(including $\sim11$\% background) were detected from all detectors combined in
two observations (also including PN data from the first observation). For
the maximum test statistics of 15.4, this implies a $3\sigma$ upper limit of
$\sim16$\%. The best limit for the PN data alone is again obtained when both
observations are combined (period search restricted to 147\,ms 100\,s to match
the resolution of the full frame data), which yields 2330 photons including 12\%
background, a peak statistics value of 17.4, and an upper limit on the pulsed fraction
of 21\% in a period range of 147\,ms to 100\,s. The upper limit for shorter periods
using the PN data from the first observation was found to be consistent with
the limit reported by \citet{Park06}, that is,$\sim45$\%.

\section{Discussion and conclusions}
A rotating neutron star will be observed to pulsate only when the hotspots
are misaligned with the rotation axis and the rotation axis itself does not
point directly to the observer. The expected pulsed fraction depends thus on
the geometrical configuration of the hotspots and on the orientation of the pulsar
with respect to the observer. This problem was discussed by \citet{Elshamouty16}
for quiescent low-mass X-ray binaries and by \citet{Suleimanov17} for other
CCOs. In particular, \citet{Suleimanov17} calculated the probability that
pulsation in this source escaped detection given the existing limits on the pulsed
fraction and hotspot size, and assuming a random location of the hotspots and
orientation of the neutron star with respect to the observer.
\citet{Suleimanov17} considered the angular distribution of the emission
emerging from the atmosphere and its propagation to the observer in the strong
gravitational field of the neutron star. These authors concluded that there is
only an $\sim8$\% chance probability that pulsations are not observed.

Here we repeated the analysis presented by \citet{Suleimanov17} for \src. We used
the limits on pulsed fraction and best-fit spectral parameters for the two-component hydrogen atmosphere obtained in the previous section as input. The
results are presented in Fig.~\ref{fig:map}, which is  similar to Fig.~14 of
\citet{Suleimanov17}. Unfortunately, the sensitivity of the observations does
not allow us to place strong constraints on the orientation of the neutron star, so that the
probabilities that the pulsations are not detected as a result of unfavorable
geometry are quite high: $\sim18.5$\% for PF$\le$16\% and $\sim30$\% for
PF$\le21$\%. Significantly longer exposures would be required to obtain stronger
constraints. We therefore conclude that the available data do not allow us to rule
out the two-component scenario based on the non-detection of the pulsations.

On the other hand, it is important to mention that \src is not the only object
for which the two-component scenario has been invoked and yet no pulsations have been
detected. As discussed by \citet{Suleimanov17}, in each individual case, the
probability of a non-detection of the pulsations as a result of unfavorable geometry is
not negligible, but it is quite unlikely that the geometry is unfavorable
for all sources. We currently know of four neutron stars that are presumably covered by carbon
envelopes: CCOs in SNRs Cas A, HESS J1731-347, G15.9+02, and G330.2+1.
None of them exhibits pulsation, and the joint probability is sufficiently low for the geometry to be unfavorable. Considering the CCOs in
HESS~J1731+347 and Cas~A, \citet{Suleimanov17} estimated a joint probability that both
these CCOs have unfavorable viewing angles toward us at
$\sim1$\%. This probability becomes even lower, about 0.3\%,\ if \src is considered. While a non-detection of the pulsations in this group of
CCOs could also be explained by some systematic process aligning the field
of the neutron star with the rotation axis, it is unclear  why the same process
would not be effective for the pulsating CCO population. We
therefore have to conclude that a separate population of CCOs
might exist whose atmospheres predominantly consist of heavier elements.

If this is indeed the case, it is important to understand the mechanisms
responsible for the enrichment of the atmosphere with heavier elements.
\citet{Chang10} suggested that in absence of accretion, diffusive nuclear
burning (DNB) of hydrogen and helium might be such a mechanism. This scenario
requires, however, that the accretion is inhibited almost completely.
\citet{Chang10} argued that a pulsar wind might be responsible for inhibiting
the accretion, but did not discuss the potential effectiveness of this mechanism.
On the other hand, \citet{Doroshenko16} suggested that
carbon in the atmosphere of the neutron star might be explained by the
accretion of material that is enriched with heavier elements during the supernova
explosion. The velocity of the expanding ejecta ($\sim
10000$\,km\,s$^{-1}$) , which almost freely expands for about one hundred years
\citep{Branch17}, is significantly higher than the typically observed kick
velocities of neutron stars ($\sim 300$\,km\,s$^{-1}$). The compact object is
therefore not expected to leave the metal-rich ejecta for several thousand
years.

In the case of HESS J1731-347 discussed by \citet{Doroshenko16}, this suggestion
is directly supported by the massive dust shell that encloses the
CCO. The dust is likely also responsible for the enhanced
absorption of the neutron star X-ray emission in this source we described above. While there is no
evidence for dust in G330.2+1.0 we discussed here, the absorption
column is similarly enhanced, which might point to additional
material around the neutron star in this case as well. \citet{Alp18} also recently
discussed such a scenario in the context of X-ray absorption in young core-collapse
SNRs, concluding that this might be the reason for
the non-detection of a CCO in SN~1987A.

\begin{acknowledgements}
VD and AS thank the Deutsches Zentrum for Luft- und Raumfahrt (DLR)
and Deutsche Forschungsgemeinschaft (DFG) for financial support. 
VS was supported by the German Research Foundation (DFG) grant WE 1312/48-1, and
he also acknowledges support by the Russian Science Foundation through grant 14-12-01287.
\end{acknowledgements}

\vspace{-0.3cm}
\bibliography{biblio}   
\vspace{-0.3cm}

\end{document}